\documentclass{easychair}

\usepackage{doc}

\usepackage{enumitem}
\usepackage{subcaption}

\title{\textbf{inbaverSim:} An OMNeT++ Model Framework for Content Centric Networking}

\author{
    Asanga Udugama
}

\institute{
  University of Bremen,
  Bremen, Germany\\
  \email{adu@comnets.uni-bremen.de}
 }

\authorrunning{Udugama}

\titlerunning{inbaverSim: CCN Model Framework}

\begin{document} 

\maketitle

\begin{abstract}
  Today's networks are used primarily to move content. To cater to this requirement 
  Information Centric Networks (ICN) were introduced. One of the main 
  architectures of ICN is Content Centric Networking (CCN) and its derivative, Named
  Data Networking (NDN). CCN is standardized at the Internet Engineering Task Force (IETF)
  and is envisaged to replace the current Internet over time. To evaluate large scale deployments
  of CCN, a model framework called the \textbf{inbaverSim} is developed in OMNeT++. This work
  presents the architecture of this model framework together with an example evaluation
  using the model framework. The code is open source and is available at GitHub.
\end{abstract}

\section{Introduction}
\label{sec:intro}

%
%

The majority use of today's networks has evolved over time to become networks for content sharing. This evolution is a problem for current host centric networks as \emph{what} content required by users have to be mapped to \emph{where} the content is located. Information Centric Networking (ICN) is a new networking paradigm that treats content as the primitive - decoupling location from identity, security and access - to retrieve content by name. Together with built-in capabilities for caching content, multi-path communications, disruption tolerance and security, ICN is able to leverage advancements in technology to address the issues associated with the mismatch between today's network use and network architecture.

Content Centric Networking (CCN) is a \emph{clean slate} ICN architecture for the future Internet, standardized at the Internet Engineering Task Force (IETF), to ultimately replace the host centric Internet protocol suite (TCP/IP). CCN, described first in \cite{ccn_09_van}, and its derivative, Named Data Networking (NDN), has been used in many other networking areas to enable content centric communications. Research ranging from flying networks \cite{icn_fly_19_lei} to sensor networks \cite{icn_sensor_18_adhatarao} and many other network types in between \cite{icn_vehicular_14_bruno, ccn_dtn_20_minamiguchi, ccn_dtn_17_islam} use CCN or NDN to enable information centric communications. 

The IETF, through the Information Centric Networking Research Group (ICNRG) has standardized CCN by approving two RFCs, namely RFC8569 \cite{RFC8569} and RFC8609 \cite{RFC8609} that describe the operation and message formats. Using these RFCs as the basis, we have developed an OMNeT++ model framework to evaluate the performance of CCNs. In this work, we describe this model framework, called the \textbf{inbaverSim}, elaborating on its constituent components and the operations.

The rest of this work is ordered in the following manner. The next section (Section~\ref{sec:sota}) provides an overview to CCN. Section~\ref{sec:models} provides the details of the model framework including the evaluation metrics. Section~\ref{sec:perf} discusses the results of a simple evaluation done using the model framework. Section~\ref{sec:summary} is a concluding summary.


\section{Content Centric Networking}
\label{sec:sota}

%
%
%

ICNs help aliviate the main problem faced by networks of today - the inability to cater to the content orientation of communications. There have been many artchitectures proposed for ICN \cite{icn_survey_12_ahlgren}. Over time, many of these architectures became obsolete and the CCN architecture together with its derivative, NDN has become the main ICN architectures.

The protocol stack as shown in \cite{ccn_09_van} consist of a number of layers, each layer providing a distinct set of operations. The fundamental focus of CCN is caching and forwarding secure content in a network. Any application, including traditional applications such as email and video streaming is able to leverage CCN to request or deliver data. Further, CCN is able to work over any transport technology that overlay network layers and a variety of link technologies. It could directly overlay any wired or wireless link technology such as Ethernet, WLAN or Bluetooth or it could overlay any traditional transport technology such as TCP of TCP/IP.

\begin{figure}[!ht]
  \centering
    \includegraphics[width=0.75\textwidth]{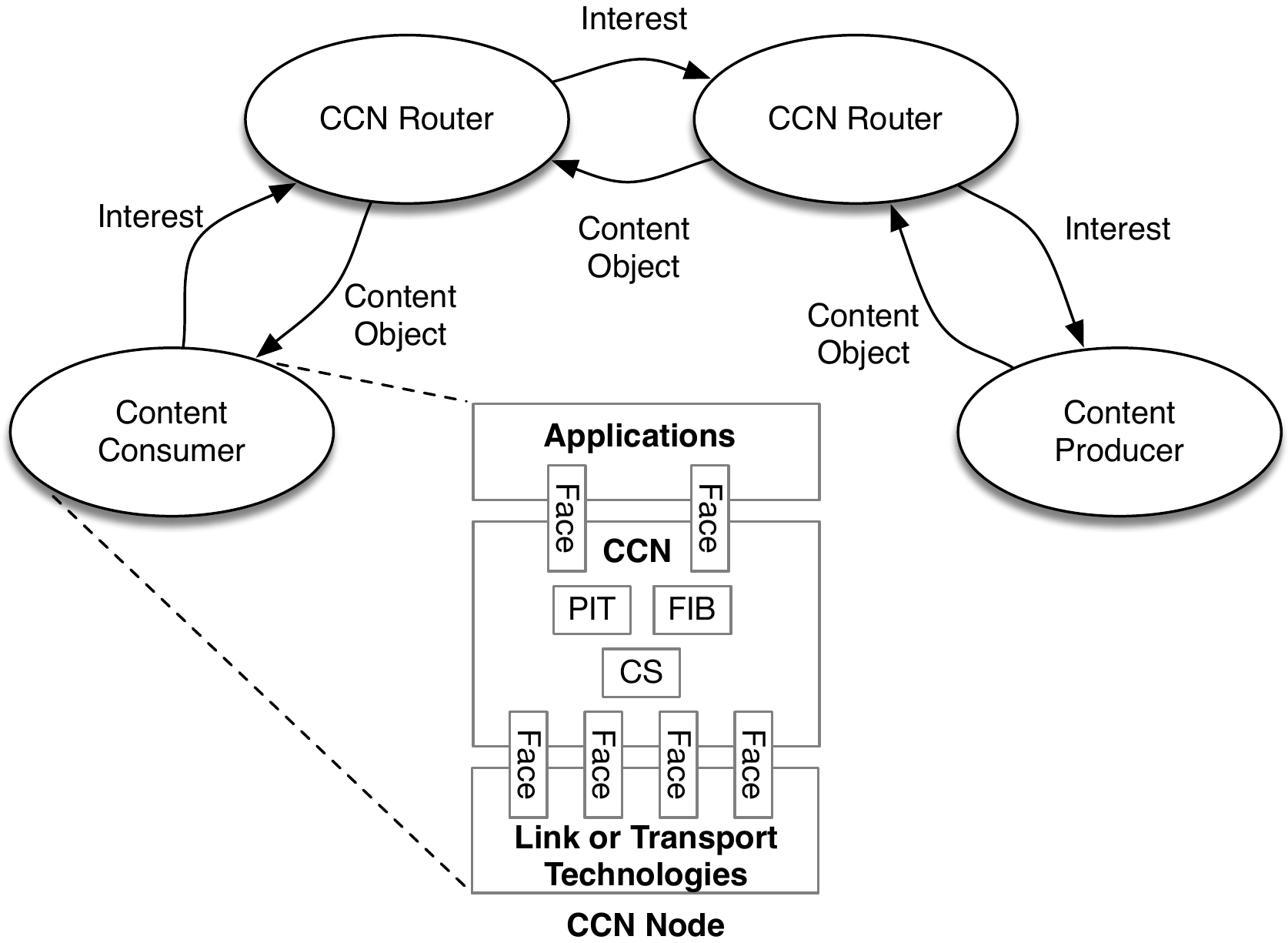}
  \caption{CCN Operation and Node Architecture}
  \label{fig:ccn-operation-and-node-arch}
\end{figure}

Figure~\ref{fig:ccn-operation-and-node-arch} shows the operation and the node architecture of CCN. The operation of CCN is based on a request-response mechanism. Every node must request for data using a message called the \textbf{Interest} and in reponse, the producer of content or a cache that receives this request responds with a \textbf{Content Object} message. The requests result in leaving a trace along the path to the location of the content and the responses find their way back to the originator of the requests using this trace, much like the bread crumbs left in the fairy tale \emph{Hansel and Gretel}.

In CCN, every content is identified uniquely by a name. The content idenified by this name may consist of multiple parts, each refered to as a segment. Every segment is identified by a number and the content name together with the segment number uniquly identifies a segment. A request (i.e., an Interest) carries this unique identifier. The response (i.e., a Content Object) carries the same identifier and the content associated with the identifier. In addition to the name, CCN messages carry other information such as security hashes and expiration times. In the case of unserved Interest messages, an \textbf{Interest Return} message is used to inform the requester.

The RFC8569 \cite{RFC8569} defines the behaviour of CCN operations while RFC8609 \cite{RFC8569} defines the message formats. Every node in a CCN network must implement the functionality described in these RFCs. CCN uses three data structures to assist in enabling caching and forwarding. The \textbf{Pending Interest Table (PIT)} holds the current requests for content, requested by a node's own applications or by other nodes, if the node is a CCN router. The \textbf{Forwarding Informartion Base (FIB)}, similar to a routing table in TCP/IP, holds a list of network attachments, refered to as \textbf{Faces} and the content that is reachable through those network attachments. The \textbf{Content Store (CS)} holds cached content, received in previous communications and that could be used in serving subsequent requests for those content.

\section{CCN Model Framework (inbaverSim)}
\label{sec:models}

%
%
%

\textbf{inbaverSim} is a model framework for OMNeT++ to simulate content centric networks. It implements the node architecture of a CCN node and a number of node models that use this node architecture. It is modular and extensible. The important aspects of the \textbf{inbaverSim} are detailed in the following sub sections.

\subsection{Protocol Layers and Model Implementations}

\begin{figure}[!ht]
  \centering
    \includegraphics[width=0.55\textwidth]{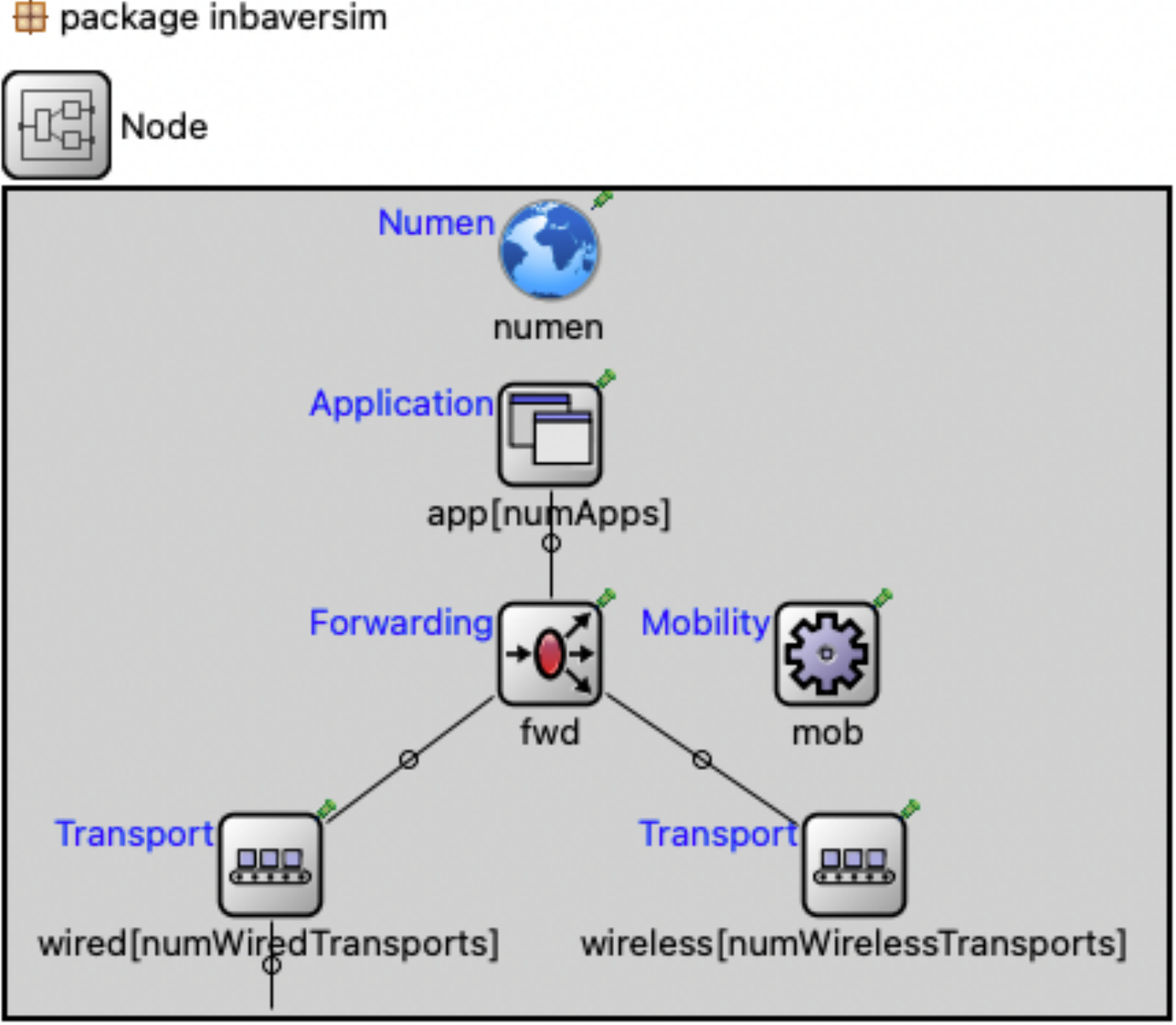}
  \caption{Protocol Stack of a CCN Node}
  \label{fig:ccn-node-model}
\end{figure}

The node architecture is based on a protocol stack that collapses the original protocol layers described in \cite{ccn_09_van} into a 3-layer protocol stack. The architecture allows the incorporation of new models at all layers of the protocol stack. Figure~\ref{fig:ccn-node-model} shows the CCN node model with the three-layer protocol implementation.

The \textbf{Application} layer contains models that mimic the behaviour of applications run in CCN nodes. They use the request-respose (i.e., Interests and Content Objects) mechanism of CCN to request and respond to requests. Currently, the following application models implemented to evaluate the performance of CCN.

\begin{itemize}[leftmargin=*]
    \item \textbf{ContentDownloadApp} - This application initiates periodic content downloads during simulations. It uses parameters such as \emph{Inter-content Download Interval} and \emph{Interest Retransmission Timeout} to control the behaviour of content downloads
    \item \textbf{ContentServerApp} - This application responds to content requests initiated by \emph{ContentDownloadApp}, by replying with content. It uses parameters such as \emph{Total Segments} and \emph{Segment Size} to determine the characteristics of the content requested.
    \item \textbf{PrefixAdvertiser} - This application is responsible for disseminating content name prefixes throughout the network much like a gateway protocol that exchange routing information. The advertised prefixes are used by CCN nodes to populate their FIBs.
    
\end{itemize}

The \textbf{Forwarding} layer contains models implementing the forwarding beviour described in CCN. Currently, the forwarding layer contains the \textbf{RFC8569Forwarder} model which implements the behaviour according to RFC8569\cite{RFC8569}. It is responsible for processing CCN messages (\emph{Interest}, \emph{Content Object} and \emph{Interest Return}) and enabling the forwarding behaviour. The behaviour consist of forwarding requests for content, returning requestes if they cannot be served, caching content and responding with content when available. To enable this behaviour, it uses the data structures \emph{PIT}, \emph{FIB} and \emph{CS}. The RFC8569 leaves out certain mechanisms required in the operation of CCN to the discretion of the implementer. An example is the use of a \emph{Cache Replacement Policy (CRP)} and the \emph{Cache Placement Strategy (CPS)} \cite{ccn_cache_20_nour}. In this implementation, we use a \emph{First-in-First-Out (FIFO)} as the CRP and \emph{Leave Copy Everywhere (LCE)} for the CPS.


In \textbf{inbaverSim}, every CCN \emph{Face} is modeled as a \textbf{Transport}. A transport is a means by which a CCN node sends and receives messages. Compared to the Open Systems Interconnection (OSI) model, in \textbf{inbaverSim}, a transport may include all layers below and including the \emph{Transport Layer}. To enable different types of link technologies, the follopwing models are implemented.

\begin{itemize}[leftmargin=*]
    \item \textbf{WiredTransport} - This model implements the end-points of a wired link between two CCN nodes for reliable communications. The user of the model is able to configure parameters such as \emph{Data Rate} and \emph{Delay} to enable different types of wired link technologies (e.g., Gigabit Ethernet).
    \item \textbf{WirelessTransport} - This model implements the end-points of wireless links. It models the basic functionality required by wireless links and is configurable with parameters such as \emph{Data Rate} and \emph{Wireless Range}. There are three possible modes of operation of this model. 
    \begin{enumerate}[leftmargin=*]
        \item \textbf{Client Mode} - This mode enables a CCN node to operate as a wireless client connecting to a wireless access point. This is similar to the station mode in WLAN. In this mode, the node only communicates with wireless access points.
        \item \textbf{Access Point Mode} - This mode acts as a wireless access point where many clients are able to connect to it. 
        \item \textbf{Direct Mode} - This mode of operation is similar to the \emph{Client Mode} above, but communicates only with other client nodes. This is similar to nodes communicating using Bluetooth or WiFi-Direct.
    \end{enumerate}
\end{itemize}

An alternative is to use similar models available in the INET model framework \cite{INET} of OMNeT++. We have noted that these models model the complete operation of wireless communications, including such areas as propagation modelling. This results in a large number of events that work against simulating large scale networks. So, we have opted for a simpler model, generating a lesser amount of simulation events.

\textbf{Mobility} of wireless nodes are modelled using the mobility modeling mechanism provided in INET model framework. INET provides a number of mobilty models ranging from trace based to synthetic models.

When the above models are used in nodes and networks, they sometimes require holding information common to multiple models. Examples are statistics and wireless information. The \textbf{Numen} and \textbf{Demiurge} models are used to hold such node and network level information, respectively.

\subsection{Node Models}

\begin{figure}[!ht]
  \centering
    \includegraphics[width=0.7\textwidth]{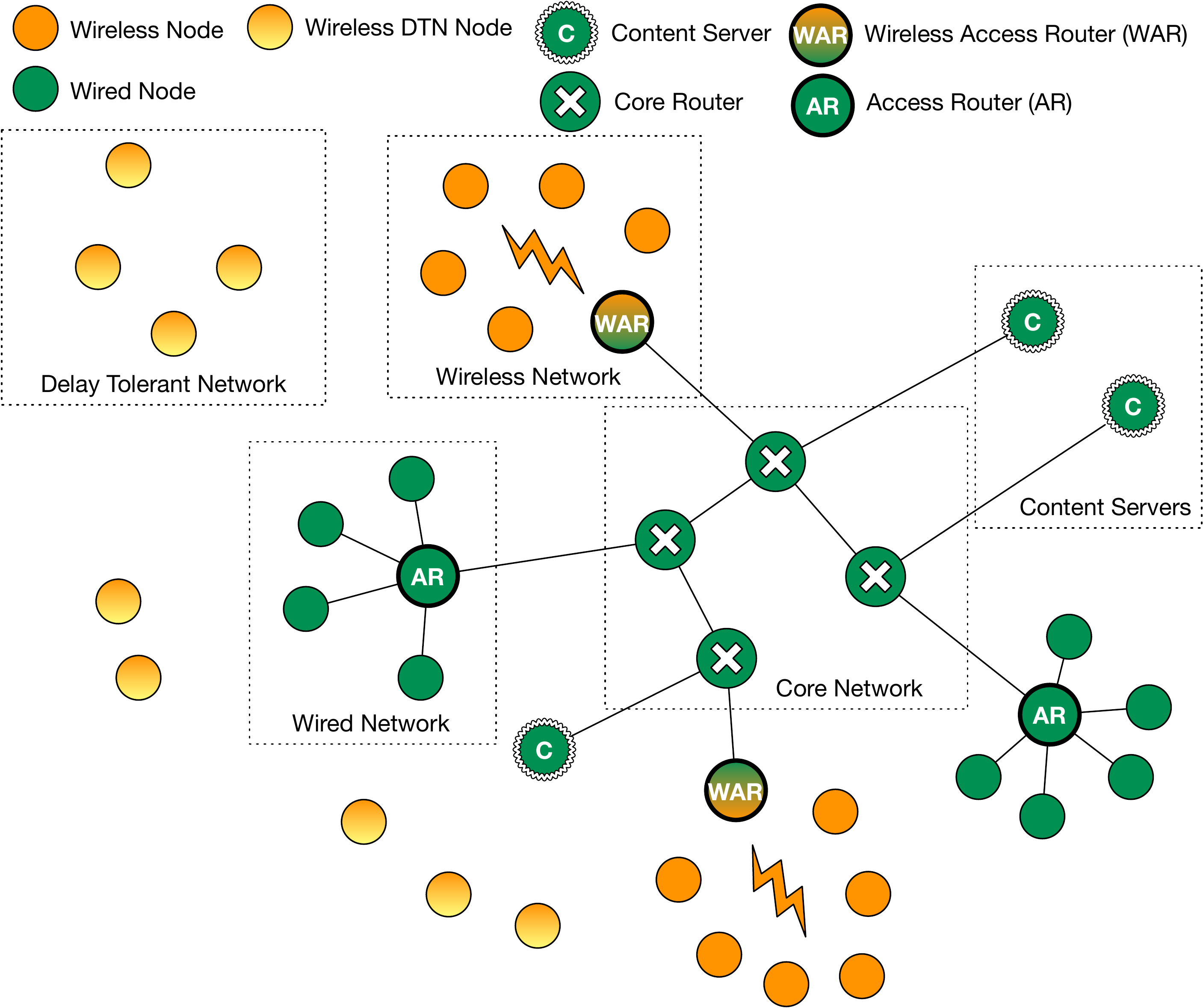}
  \caption{Network Topology and Node Models in CCN}
  \label{fig:ccn-network-topology}
\end{figure}

The above described models of the three-layer protocol stack form the basis for every CCN node. Though the forwarding functionality is the same in each CCN node, based on the primary purpose of a CCN node, there are different types of CCN nodes. Figure~\ref{fig:ccn-network-topology} shows these different CCN nodel models and the topologies of the networks they are deployed in.

The differentiation of the node models occur based on the applications deployed, transports used and the configurations of models. An example is the data rate of a wireless transport which can be set according to the wireless technology being considered. These node models are described below.

\begin{itemize}[leftmargin=*]
    \item \textbf{Wireless Node} - This is a node model that is equipped with a wireless transport to connect to a Wireless Access Router (WAR). Once such a node is in the wireless range of a WAR, it will send and receive CCN messages to and from the WAR.
    
    \item \textbf{Wireless Access Router (WAR)} - This node model is equipped with a wireless transport and a configurable number of wired transports. Its operation is similar to a WLAN access point.
    
    \item \textbf{Wireless DTN Node} - This node model is equipped with two wireless transports, one for direct communications and the other connecting to a WAR. Depending on the nodes in its wireless range, it may communicate over one or both transports.
    
    \item \textbf{Wired Node} - This is a node model equiped with a wired transport to connect with an Access Router (AR).
    
    \item \textbf{Access Router (AR)} - This is a node model equiped with multiple wired transports to connect with multiple wired client nodes or other ARs. It is similar in functionality to a router in the Internet. 

    \item \textbf{Content Server} - This model is equipped with a wired transport and is deployed with the \emph{ContentServerApp} to serve content requests arriving from other CCN nodes. These nodes host content.
    
    \item \textbf{Core Router} - This is a node model that is equipped with multiple wired transports to forward Interests in the direction of where the content are hosted. The content are hosted in Content Servers and the FIBs of the Content Routers are populated based on the content prefixes hosted in these Content Servers. The \emph{PrefixAdvertise} applications are deployed in these node models to disseminate content prefixes. They are operate similar to the core routers hosted in the backbone of the Internet.
    
\end{itemize}

These are the basic node models. But, new node models with different combinations of transports are possible due to the extensibility of the model framework. For example, a new \emph{Wired CCN Node} with multiple wired transports can be created to evaluate the performance of multi-path content deliveries in CCN.

%
%
%
%
%
%
%
%
%

\subsection{Evaluation Metrics}

Each protocol model in \textbf{inbaverSim} generates different metrics based on its functionality. These metrics may be collected at the \emph{protocol level}, \emph{node level} or at the \emph{network level}. Listed below are a selected set of these evaluation metrics. An exhaustive list of the all metrics are available at the Github repository.

\begin{itemize}[leftmargin=*]
    \item \textbf{Average Cache Hit/Miss Ratio} These metrics computes a ratio of all cache hits or misses to the the total cache requests made. Usually a higher hit ratio indicates the availability of content being requested in most caches. A higher miss ratios may be due to multiple reasons. For example, poor hit ratios may be due to failures in proper cache provisioning.

    \item \textbf{Average Content/Segment Download Duration} An Interest sent out ideally brings back a segment of data of a content. If the caches are highly fragmented, then the indivudual segment delays may fluctuate rapidly. Similarly, if the caches are not populated with the requested content, then the Interest must travel all the way to the content producer. Factors such as these may influence delays. These metrics gives an idea of the latency of content retrievals. 

    \item \textbf{Total Interest Retransmissions Bytes Received/Sent} Interets bring back segments. Ideally, only one Interest is required to be issued to receive the corresponding segment of data. But sometimes, due to the movement of nodes, Interests ay need to be re-issued thereby increasing the number of sent Interests. These metrics gives an idea of the retransmission overhead in bytes incurred when requesting content.

    \item \textbf{Average PIT Entry Count} The number of entries in the PIT at any given time indicates how well the Interests are bringing back content. A steady average of this metric indicates a good content flow. Fluctuations may indicate problems such as cache fragmentations.

\end{itemize}

\section{Performance Evaluation}
\label{sec:perf}

%

In this section, we present a simple evaluation of a CCN network using \textbf{inbaverSim}. The objective of the evaluation is to understand the cache behaviour when varying the diversity of the content catalog in the network. Diversity refers to the scale of the variety of content present in the network for users to download. The lower the diversity, the lesser the choice is for a user to download a unique content, not downloded before. Higher the diversity, the higher the choice is to download a unique content.


\begin{figure}[!ht]
  \centering
    \includegraphics[width=0.6\textwidth]{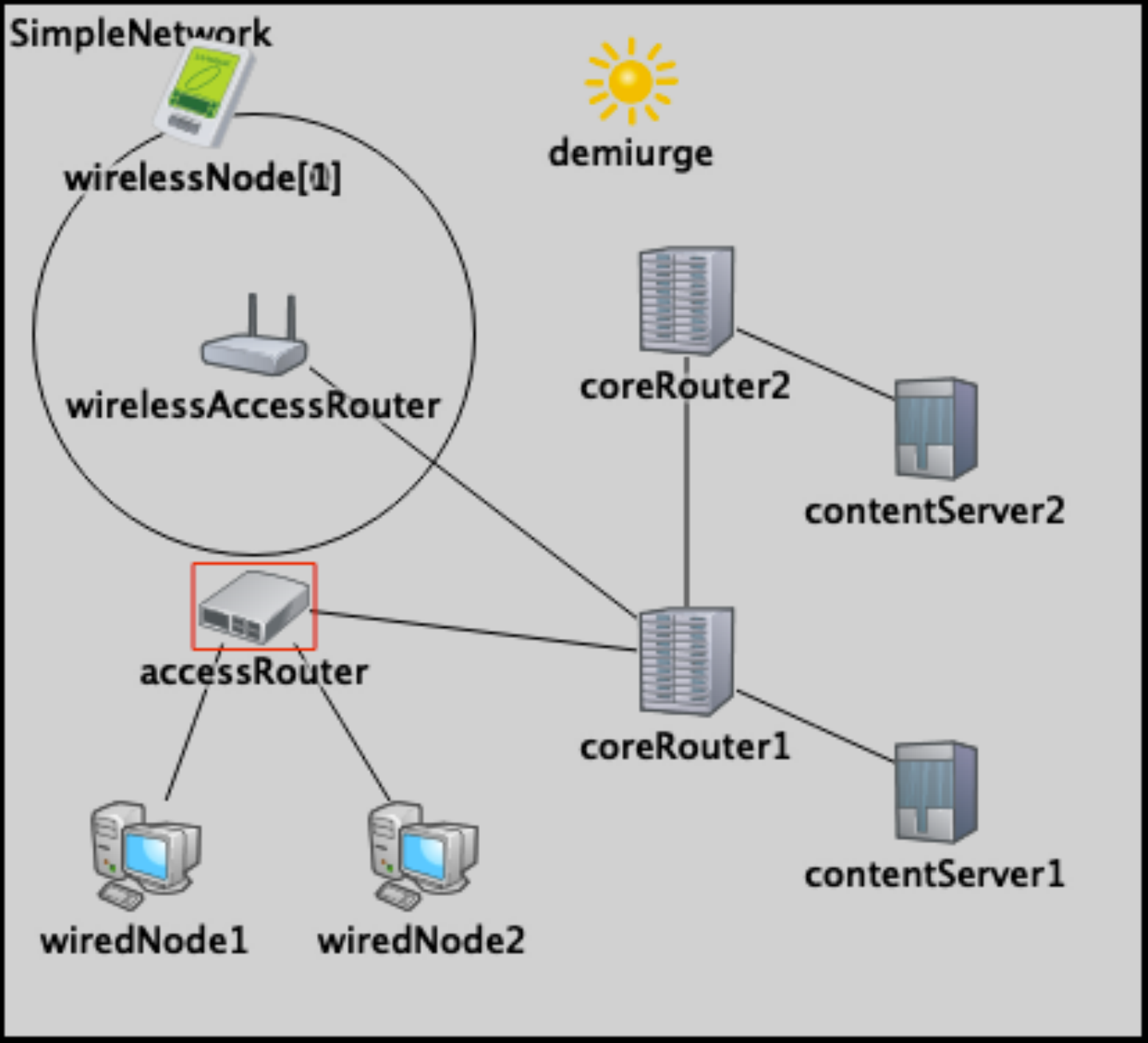}
  \caption{CCN Network used for the Evaluation}
  \label{fig:simple-ccn-network}
\end{figure}

Figure~\ref{fig:simple-ccn-network} shows the network created to evaluate the effects of content diversity. The evaluation scenario consist of four CCN clients (two wireless and two wired) retrieving content from two content servers. Mobility of the wireless nodes are realized using the \emph{RandomWaypointMobility} model of the INET model framework. The variation of content diversity is handled by changing the content catalog sizes of content at content servers. Three levels of content catalog variations are considered, referenced as \emph{High}, \emph{Medium} and \emph{Low}. The simulations are run for a 48-hour period.


\begin{figure*}[!ht]
    \centering
    \begin{subfigure}[b]{0.7\textwidth}
        \includegraphics[width=\textwidth]{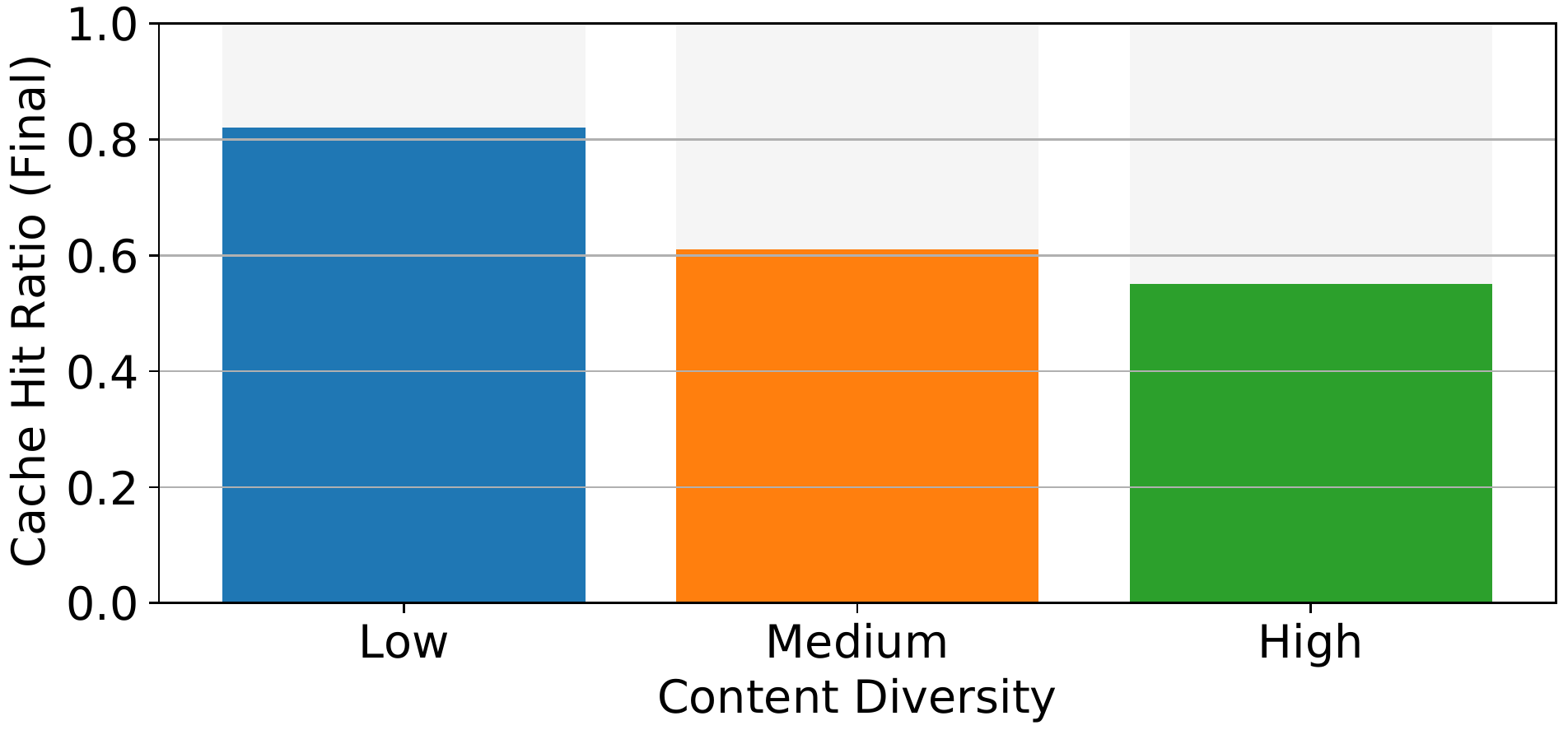}
        \caption{Cache Hit Ratios after 48 hours}
        \label{fig:cache-hit-miss-ratio-bar}
    \end{subfigure}
    \begin{subfigure}[b]{0.7\textwidth}
        \includegraphics[width=\textwidth]{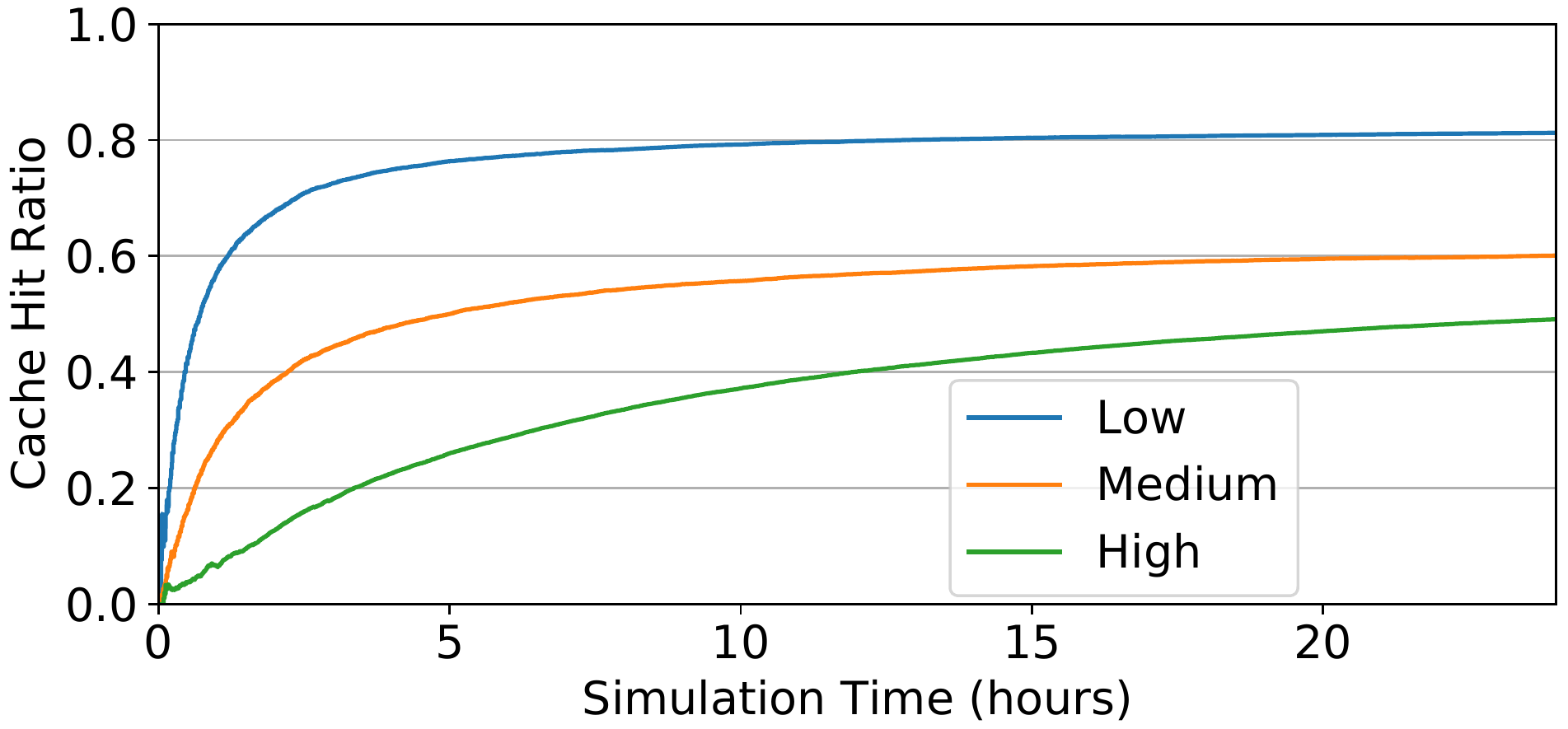}
        \caption{Progression of the Cache Hit Ratios for the first 24 hours}
        \label{fig:cache-hit-ratio-graph}
    \end{subfigure}
    \caption{Cache Hit Ratios when varying content catalog sizes in the network}
    \label{fig:cache-hit-ratio}
\end{figure*}

Figure~\ref{fig:cache-hit-miss-ratio-bar} shows the network-wide \emph{Cache Hit Ratios} achieved for each of the catalog types. The results show that when the diversity increases, the cache hit ratio degrades. When the catalog is highly diverse (\emph{High} in graph), due to the randomness of requests for content, the requests become equally diverse. This results in most requests for content reaching the content servers as the nearby caches have not seen the content yet. Therefore, along the way, all intermediate CCN nodes such as the core routers report a high amount of cache misses. As a result, the network-wide cache hit ratios degrade as seen from the poor hit ratio in the graph. The logic is vice versa for content catalogs that are less diverse. Lesser the variety of requests, more the likeliness of finding the content cached in a nearby cache, as seen from the cache hit ratio for \emph{Low}.

Figure~\ref{fig:cache-hit-ratio-graph} shows a comparison of the progression of the \emph{Cache Hit Ratio} for the three catalog sizes for a 24-hour period. This graph confirms the explanation given above regarding the behaviour based on the content catalog diversity (e.g., highly diverse catalogs result in lower cache hit ratios). The initial low readings for all catalog sizes are due to most requests for content reaching the content servers. The progression shows that the ratios are gradually reaching their steady states.

\section{Summary and Future Work}
\label{sec:summary}

%
%

In this work, an OMNeT++ model framework for CCN called \textbf{inbaverSim} is presented. This model framework implements the CCN functionality as described in RFC8569 and RFC8609. The model framework has a modular and extensible architecture. This architecture enables uncomplicated incorporation of new or extended models to the framework.

The behaviour described in RFC8569 provides the basic functionality required to operate in traditional networks (infrastructure-based networks). CCN is an ideal candidate for many types of networks to operate at the forwarding/routing layers of the protocol stack. But the behaviour described in the RFCs is inefficient due to the nature of many networks. An example is intermittently connected networks such as Delay Tolerant Networks (DTNs) \cite{ccn_dtn_20_minamiguchi} or Opportunistic Network (OppNets) \cite{ccn_dtn_17_islam}. Therefore, in the future, \textbf{inbaverSim} is planned to be extended to operate in other networks such as OppNets and Vehicular Networks \cite{icn_vehicular_14_bruno}.

The framework code is open source and is available at GitHub\footnote{https://github.com/ComNets-Bremen/inbaverSim}.

\label{sec:bib}
\bibliographystyle{plain}
\bibliography{paper}

\begin{thebibliography}{10}

\bibitem{icn_sensor_18_adhatarao}
Sripriya~Srikant Adhatarao, Mayutan Arumaithurai, Dirk Kutscher, and Xiaoming
  Fu.
\newblock Isi: Integrate sensor networks to internet with icn.
\newblock {\em IEEE Internet of Things Journal}, 5(2):491--499, 2018.

\bibitem{icn_survey_12_ahlgren}
Bengt Ahlgren, Christian Dannewitz, Claudio Imbrenda, Dirk Kutscher, and Borje
  Ohlman.
\newblock A survey of information-centric networking.
\newblock {\em IEEE Communications Magazine}, 50(7):26--36, 2012.

\bibitem{icn_vehicular_14_bruno}
Federico Bruno, Matteo Cesana, Mario Gerla, Giulia Mauri, and Giacomo
  Verticale.
\newblock Optimal content placement in icn vehicular networks.
\newblock In {\em 2014 International Conference and Workshop on the Network of
  the Future (NOF)}, pages 1--5, 2014.

\bibitem{ccn_dtn_17_islam}
Hasan M.~A. Islam, Dimitris Chatzopoulos, Dmitrij Lagutin, Pan Hui, and Antti
  Ylä-Jääski.
\newblock Boosting the performance of content centric networking using delay
  tolerant networking mechanisms.
\newblock {\em IEEE Access}, 5:23858--23870, 2017.

\bibitem{ccn_09_van}
Van Jacobson, Diana~K. Smetters, James~D. Thornton, Michael~F. Plass,
  Nicholas~H. Briggs, and Rebecca~L. Braynard.
\newblock Networking named content.
\newblock In {\em Proceedings of the 5th International Conference on Emerging
  Networking Experiments and Technologies}, CoNEXT '09, page 1–12, New York,
  NY, USA, 2009. Association for Computing Machinery.

\bibitem{icn_fly_19_lei}
Kai Lei, Qichao Zhang, Junjun Lou, Bo~Bai, and Kuai Xu.
\newblock Securing icn-based uav ad hoc networks with blockchain.
\newblock {\em IEEE Communications Magazine}, 57(6):26--32, 2019.

\bibitem{ccn_dtn_20_minamiguchi}
Chuta Minamiguchi, Ryo Nakamura, and Hiroyuki Ohsaki.
\newblock Comparative analysis of content routing strategies in
  information-centric delay-tolerant networking.
\newblock In {\em 2020 International Conference on Information Networking
  (ICOIN)}, pages 778--783, 2020.

\bibitem{RFC8609}
M.~Mosko, I.~Solis, and C.~Wood.
\newblock Content-centric networking (ccnx) messages in tlv format.
\newblock RFC 8609, RFC Editor, July 2019.

\bibitem{RFC8569}
M.~Mosko, I.~Solis, and C.~Wood.
\newblock Content-centric networking (ccnx) semantics.
\newblock RFC 8569, RFC Editor, July 2019.

\bibitem{ccn_cache_20_nour}
Boubakr Nour, Hakima Khelifi, Hassine Moungla, Rasheed Hussain, and Nadra
  Guizani.
\newblock A distributed cache placement scheme for large-scale
  information-centric networking.
\newblock {\em IEEE Network}, 34(6):126--132, 2020.

\bibitem{INET}
OMNeT~Core Team.
\newblock {OMNeT INET Model Framework}.
\newblock \url{https://inet.omnetpp.org}, 2021.
\newblock [Online; accessed 25-July-2021].

\end{thebibliography}


\end{document}